\documentclass{article}
\usepackage{arxiv}
\usepackage[utf8]{inputenc}
\usepackage[T1]{fontenc}
\usepackage{lmodern}
\usepackage{xcolor}
\usepackage{amsmath,amssymb,amsthm}
\usepackage[numbers,square,comma,sort]{natbib}
\usepackage{hyperref}
\usepackage{url}
\usepackage{doi}
\usepackage{booktabs}
\usepackage{graphicx}
\usepackage{microtype}
\usepackage{enumitem}
\setlist{leftmargin=*,labelindent=0pt,labelsep=0.5em}
\hypersetup{colorlinks=true,linkcolor=blue!50!black,citecolor=blue!50!black,urlcolor=blue!50!black,
  pdftitle={The Scope Grammar of Natural-Language Rules: Counting and Enumerating Readings modulo Logical Equivalence},
  pdfauthor={Sherzod Turaev}}

\newcommand{\EXC}{\mathbin{\smallsetminus}}
\newcommand{\sem}[1]{[\![#1]\!]}
\newcommand{\BB}{\mathbb{B}}

\theoremstyle{plain}
\newtheorem{theorem}{Theorem}[section]
\newtheorem{lemma}[theorem]{Lemma}
\newtheorem{proposition}[theorem]{Proposition}
\newtheorem{corollary}[theorem]{Corollary}
\newtheorem{conjecture}[theorem]{Conjecture}
\theoremstyle{definition}
\newtheorem{definition}[theorem]{Definition}
\theoremstyle{remark}
\newtheorem{remark}[theorem]{Remark}

\title{The Scope Grammar of Natural-Language Rules:\\ Counting and Enumerating Readings modulo Logical Equivalence}

\author{
  Sherzod Turaev\thanks{Corresponding author: \texttt{sherzod@uaeu.ac.ae}. ORCID 0000-0001-6661-8469.}\\
  Department of Computer Science and Software Engineering\\
  College of Computing and Artificial Intelligence\\
  United Arab Emirates University, Al Ain, United Arab Emirates\\
  \texttt{sherzod@uaeu.ac.ae}
}
\date{}

\begin{document}
\maketitle

\begin{abstract}
Automated reasoning can certify a rule only once that rule has been written in logic, yet the rules for which such certification is now wanted, from authorization and business policies to clinical eligibility criteria and academic regulations, exist first as text written for people. Such a text fixes which atomic conditions and which connectives occur, but it leaves open how those connectives are bracketed, and each bracketing is a \emph{different} logical reading that an autoformalizer must either resolve or expose. We investigate the combinatorial object behind this choice, the \emph{scope grammar}, a small and deliberately ambiguous context-free grammar whose derivation trees are the readings of a connective skeleton and whose quotient by logical equivalence is what matters downstream. A connective skeleton with $k$ binary connectives and no negation has $C_k$ derivation trees, the $k$-th Catalan number, but far fewer logically distinct readings, because associativity and the laws of the Boolean algebra identify many of the trees. For the positive fragment, repetition-free and negation-free, we make the gap exact: a reading computes a read-once function, and the uniqueness of the read-once normal form lets us characterize the realizable normal forms by the labels of lowest common ancestors and count them with a two-label recurrence over the operator word, which we verify against exhaustive enumeration for every pattern up to eleven atoms. The count is invariant under reversal and under the $\wedge/\vee$ duality, and is therefore a function of the composition of operator runs. We prove that a positive skeleton is logically ambiguous if and only if it uses both $\wedge$ and $\vee$, a one-pass test, and that the number of reading classes is at least $2^{\lfloor n/3\rfloor}$ on $n$ atoms. For every pattern up to fifteen atoms the maximally alternating pattern is the worst case, and we determine its class count exactly: its generating function is algebraic of degree three, the count is the 3-Schr\"oder sequence (OEIS A137842), and its growth base is $\varphi^{5/2}=\sqrt{(11+5\sqrt5)/2}\approx3.330$, the five-halves power of the golden ratio and strictly below the Catalan base $4$. The maximality itself we reduce to an inequality on increments, verified to sixteen atoms, and we explain why local exchange arguments cannot establish it. We then treat negation, exceptions, and background hyponymy axioms, where normal-form uniqueness still decides equivalence though the counting law does not transfer, and we locate the work among the ambiguity of context-free grammars, read-once and series-parallel enumeration, and the underspecification of quantifier scope in computational semantics.
\end{abstract}


\section{Introduction}\label{sec:intro}

Automated reasoning can give guarantees about a rule only once that rule has been written in logic, and in the settings where such guarantees are now wanted, from authorization and business policies to clinical eligibility criteria and academic regulations, the rules exist first as text that people wrote for other people. Translating them is the task of \emph{autoformalization}, which large language models now perform fluently \cite{wu2022,cosler2023}, and on which industrial systems have begun to depend, for example in the Automated Reasoning checks of Amazon Bedrock Guardrails.\footnote{Amazon Bedrock Guardrails, Automated Reasoning checks: \url{https://aws.amazon.com/bedrock/guardrails/}.} The value of every downstream proof is bounded by the fidelity of this translation, and a principal way fidelity fails is \emph{scope ambiguity}: a sentence fixes which atomic conditions and which connectives occur, yet leaves open how the conjunctions, disjunctions, negations, and exceptions are bracketed, so that ``no $A$ or $B$ within thirty days'' has several logical readings that disagree about who is covered. A measurement study of this phenomenon, reported separately, finds that for roughly one clinical eligibility criterion in six a plausibility judge retains two or more grammar-generated readings as readings of the text \cite{p01}.

It is therefore natural to separate a translation into a stage that commits to \emph{what} the atoms and connectives are, which a language model does well, and a stage that enumerates \emph{how} they may be bracketed, which a small context-free grammar can do exhaustively. The grammar in question,
\[
E\;\to\; E\wedge E \mid E\vee E \mid E\EXC E \mid \neg E \mid a,\quad a\in A,
\]
which we call the \emph{scope grammar}, is ambiguous by design, and its derivation trees over a fixed sequence of atoms and connectives, the \emph{skeleton}, are exactly the candidate readings. What matters downstream is not the trees themselves but their quotient by logical equivalence, because two bracketings that denote the same Boolean function are the same rule. Consider the skeleton $a\wedge b\vee c\wedge d$: it has $C_3=5$ bracketings, but only four are logically distinct, because $(a\wedge(b\vee c))\wedge d$ and $a\wedge((b\vee c)\wedge d)$ are the same function by associativity of conjunction (Figure~\ref{fig:nf}). The questions this raises are how many trees a skeleton admits, how many equivalence classes, which syntactic restrictions reduce the class count, and how cheaply the classes can be decided and enumerated.

We answer these questions for the positive fragment, repetition-free and negation-free, where a complete theory is available, and we treat negation and background axioms in part. The object sits at the meeting of three established lines of work, the ambiguity of context-free grammars, the theory of read-once Boolean functions, and the underspecification of scope in computational semantics, which we survey in Section~\ref{sec:related}. Though each of these lines illuminates a facet of the object, none of them counts the Boolean quotient of the bracketings of a single skeleton, which is precisely the quantity an autoformalizer must control. In the current research, we make the following contributions.
\begin{enumerate}[nosep,topsep=2pt]
\item We give a dynamic program that enumerates the reading classes directly over Boolean functions, under span-local admissibility constraints, with a correctness proof and a complexity bound (Sections~\ref{sec:adm} and \ref{sec:enum}).
\item For a positive skeleton we prove, through the uniqueness of the read-once normal form, a characterization of the realizable normal forms by lowest common ancestors (Lemma~\ref{lem:lca}) and, from it, an exact two-label counting recurrence (Theorem~\ref{thm:count}); the count is invariant under reversal and the $\wedge/\vee$ duality, hence a function of the composition of operator runs (Proposition~\ref{prop:sym}), and it agrees with exhaustive enumeration for every pattern up to eleven atoms (Section~\ref{sec:count}).
\item We prove that a positive skeleton is logically ambiguous if and only if it uses both $\wedge$ and $\vee$, a one-pass test (Theorem~\ref{thm:crit}), and that the number of classes is at least $2^{\lfloor n/3\rfloor}$ (Corollary~\ref{cor:lb}); we verify for every pattern up to fifteen atoms that the maximally alternating pattern is the worst case (Proposition~\ref{prop:max}), we show that its count has a cubic generating function, identify it with the 3-Schr\"oder sequence A137842, and obtain its growth base $\varphi^{5/2}=\sqrt{(11+5\sqrt5)/2}\approx3.330$, strictly below the Catalan base $4$ (Theorem~\ref{thm:growth}); and we reduce the maximality conjecture, through a canonical-bracketing recurrence (Proposition~\ref{prop:append}), to an inequality on increments that we verify to sixteen atoms and show to be beyond local exchange arguments.
\item We extend the equivalence test to negation, exceptions, and background hyponymy axioms, where normal-form uniqueness still decides equivalence but the counting law does not transfer, and we relate the axiomatized ambiguity test to Boolean entailment (Section~\ref{sec:beyond}).
\end{enumerate}
The paper is organized as follows. Section~\ref{sec:related} surveys related work and Section~\ref{sec:prelim} fixes the preliminaries. Section~\ref{sec:grammar} defines the scope grammar and its readings, Section~\ref{sec:adm} the admissibility constraints, and Section~\ref{sec:enum} the enumeration algorithm. Section~\ref{sec:count} is the counting theory for the positive fragment, Section~\ref{sec:beyond} treats negation, exceptions, and background axioms, and Section~\ref{sec:concl} concludes with open problems.

\section{Related work}\label{sec:related}

\paragraph{Ambiguity of context-free grammars.}
That a context-free grammar may assign many derivation trees to one string, and that some context-free languages are inherently ambiguous, are classical \cite{ginsburgullian,harrison}, and the \emph{degree of ambiguity} as a function of the word length has been classified down to the distinction between polynomial and exponential growth \cite{wich}. The scope grammar is deliberately and exponentially ambiguous; our concern is not the ambiguity of the grammar as a language but the Boolean quotient of the derivation trees of a single string, a question this literature does not raise. Regulated rewriting \cite{dassowpaun,dassowturaev} supplies the vocabulary for the admissibility constraints of Section~\ref{sec:adm}, which restrict the legal derivations of a fixed string rather than the generated language.

\paragraph{Read-once functions and series-parallel networks.}
A Boolean function expressible with each variable used once is \emph{read-once}; such functions have a unique formula up to associativity and commutativity, equivalently a unique cotree, and their co-occurrence graphs are exactly the cographs \cite{hayes1975,gurvich1977,mundici,karchmer,corneil,golumbicgurvich}. We use this uniqueness to turn the counting of reading classes into the counting of realizable normal forms. Positive read-once functions coincide with two-terminal series-parallel networks, whose enumeration is classical \cite{riordanshannon}; our count differs in that it fixes the left-to-right order of the leaves and the operator between each adjacent pair. The closest work in spirit is the enumeration of truth-table statistics over the bracketings of a single operator string: Cameron and Yildiz \cite{cameronyildiz}, and Yildiz in follow-up work \cite{yildiz}, count the true and false entries summed over the truth tables of all bracketings of $x_1\Rightarrow\cdots\Rightarrow x_n$, obtaining a recurrence and a limiting proportion of false rows. Implication is neither associative nor read-once, and their quantity is a sum over the rows of all bracketings rather than a count of distinct functions, so their combinatorics and ours diverge; what the two share is the object, the truth tables of the bracketings of one operator string. For one word, the maximally alternating one, the class count itself turns out to be on record: the sequence $1,2,4,10,24,66,172,\dots$ is entry A137842 of the OEIS \cite{oeisA137842}, the interleaving of the \emph{3-Schr\"oder numbers} A027307 and A032349 \cite{oeisA027307,oeisA032349}, which count lattice paths with steps $(2,1)$, $(1,2)$, $(1,-1)$ and go back to a Monthly problem of Deutsch \cite{deutsch2000} (see also Yang and Jiang \cite{yangjiang2021}), and Burstein recorded in 2023, as a comment on these entries, that they count the distinct sets obtainable by parenthesizing an alternating expression of unions and intersections, which is the reading-class count of that word. We learned of this after deriving Theorem~\ref{thm:growth}; our derivation is independent and identifies the sequence through its generating function, and the theory for arbitrary operator words (the characterization, the recurrence, the symmetries, the criterion, and the lower bound of Section~\ref{sec:count}) has, as far as we can determine, no precedent.

\paragraph{Underspecification of scope in computational semantics.}
The counting and compact representation of readings is well developed for \emph{quantifier} scope, where a single underspecified description stands for a family of scopings: the generation of quantifier scopings \cite{hobbs1987}, hole semantics and Minimal Recursion Semantics \cite{copestake2005}, and dominance constraints \cite{niehren}. That tradition, moreover, does quotient readings by logical equivalence: redundancy elimination removes logically equivalent scopings from an underspecified chart \cite{kollerthater06}, and the set of readings is represented and enumerated as the language of a regular tree grammar \cite{kollerregnerithater08}. Our work is the connective-scope counterpart. Two differences make the Boolean quotient sharper here than the rewrite-rule equivalence used for quantifiers: the objects are bracketings of $\wedge,\vee,\neg$ rather than orderings of quantifiers, and equivalence is exact Boolean equality, for which the read-once normal form gives a complete invariant and a closed counting law rather than an elimination procedure. Scope ambiguity is also where current language models are known to be unreliable \cite{kamath}, which is the practical reason to enumerate rather than resolve it; the empirical side of that argument, for clinical eligibility criteria, is the companion study \cite{p01}. The same concern with ambiguity in rules written for people drives neurosymbolic auditing of software requirements \cite{halleiers} and the older linguistic analysis of ambiguity in requirements and contracts \cite{berry2003}.

\section{Preliminaries}\label{sec:prelim}

Let $A$ be a finite set of \emph{atoms} with $|A|=m$, and let $\BB_m$ be the set of Boolean functions $\{0,1\}^A\to\{0,1\}$. For $\Gamma\subseteq\BB_m$ a set of \emph{background axioms}, two functions $f,g$ are \emph{$\Gamma$-equivalent} when $\Gamma\models f\leftrightarrow g$, that is $f$ and $g$ agree on every model of $\Gamma$; for $\Gamma=\emptyset$ this is equality of functions. A function \emph{depends on} (has as \emph{essential}) a variable $a$ if some assignment changes value when $a$ is flipped; it is \emph{monotone} if flipping an input from $0$ to $1$ never changes the value from $1$ to $0$.

A function is \emph{read-once} if it has a formula over $\{\wedge,\vee,\neg\}$ in which every variable occurs once, and \emph{positive read-once} if such a formula exists over $\{\wedge,\vee\}$. We use the following classical fact; it is the one external result the counting rests on.

\begin{proposition}[Read-once normal form, see \cite{golumbicgurvich}]\label{prop:readonce}
A positive read-once function has a read-once formula over $\{\wedge,\vee\}$ that is unique up to the associativity and commutativity of the two operators. Equivalently, contracting every maximal chain of one operator into a single multi-ary gate yields a tree with strictly alternating gate labels, the \emph{normal form}, which is the unique cotree of the function's cograph.
\end{proposition}

Immediately, every variable of a read-once formula is essential: a formula in which each variable occurs once depends on each, since fixing the other inputs leaves a formula in that one variable that is not constant. We use this fact repeatedly below.

A \emph{bracketing} of a sequence of $n$ operands is a full binary tree with the operands as leaves in order; there are $C_{n-1}=\frac{1}{n}\binom{2n-2}{n-1}$ of them, the $(n{-}1)$-th Catalan number \cite{stanley}. By an \emph{ordered tree} on $a_1,\dots,a_n$ we mean a rooted tree whose leaves are these atoms in left-to-right order. We write $w=o_1\cdots o_{n-1}$ for an operator word, $o_k$ for the connective between $a_k$ and $a_{k+1}$, and, in an ordered tree, $\mathrm{lca}(k)$ for the lowest common ancestor of $a_k$ and $a_{k+1}$.

\section{The scope grammar and its readings}\label{sec:grammar}

\begin{definition}[Skeleton]
A \emph{skeleton} is a word $w=w_1\cdots w_N$ of length $N$ over $\Sigma=A\cup\{\wedge,\vee,\EXC,\neg,\|\}$, where $\wedge,\vee,\EXC$ are binary connectives written between their operands, $\neg$ is a prefix connective written at the position of the negation cue, and $\|$ marks a sentence or list-item boundary. A skeleton is \emph{repetition-free} when each atom of $A$ occurs at most once in $w$. We reserve $n$ for the number of atoms of $w$ and $N$ for its length as a string.
\end{definition}

\begin{definition}[Scope grammar]
The \emph{scope grammar} is $G=(\{S,E\},\Sigma,P,S)$ with
\[
\begin{gathered}S\to S\,\|\,S \mid E,\\ E\to E\wedge E \mid E\vee E \mid E\EXC E \mid \neg E \mid a \quad (a\in A).\end{gathered}
\]
A \emph{reading} of $w$ is a derivation tree of $w$ in $G$; the set of readings is $T(w)$. A boundary is interpreted as conjunction, and no connective inside a sentence takes an operand across a boundary.
\end{definition}

The meaning $\sem{t}\in\BB_m$ of a reading is defined compositionally by $\sem{a}=$ the projection on $a$, $\sem{t_1\wedge t_2}=\sem{t_1}\wedge\sem{t_2}$, $\sem{t_1\vee t_2}=\sem{t_1}\vee\sem{t_2}$, $\sem{t_1\EXC t_2}=\sem{t_1}\wedge\neg\sem{t_2}$, $\sem{\neg t}=\neg\sem{t}$, and $\sem{t_1\|t_2}=\sem{t_1}\wedge\sem{t_2}$. The exception connective is not a syntactic abbreviation of ``$\wedge\neg$'': in $t_1\EXC t_2$ the negation covers exactly the constituent $t_2$, whereas a $\neg$ token may scope over several spans, which is itself a source of ambiguity (Section~\ref{sec:beyond}).

\begin{definition}[Reading classes]
For a set $\Gamma$ of background axioms over $A$ (for example $e\to h$ when the atom $e$ names a kind of the atom $h$), two readings are \emph{$\Gamma$-equivalent}, written $t_1\equiv_\Gamma t_2$, when $\sem{t_1}$ and $\sem{t_2}$ are $\Gamma$-equivalent. The set of \emph{reading classes} is $R_\Gamma(w)=T(w)/{\equiv_\Gamma}$, written $R(w)$ when $\Gamma=\emptyset$. The skeleton $w$ is \emph{logically ambiguous} when $|R_\Gamma(w)|\ge2$.
\end{definition}

\begin{figure}[t]
\centering
\includegraphics[width=\textwidth]{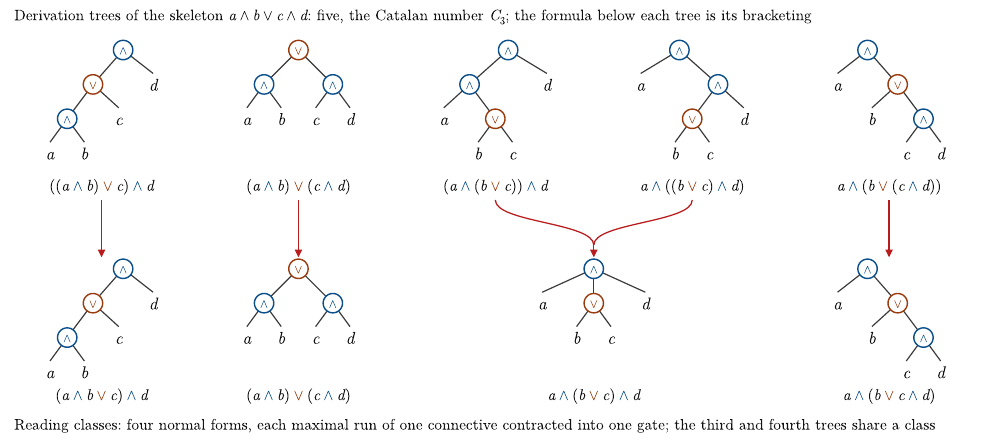}
\caption{The skeleton $a\wedge b\vee c\wedge d$ has $C_3=5$ derivation trees (top) but four reading classes (bottom): the third and fourth trees differ only in the association of the two conjunctions and contract to the same normal form, in which the root conjunction has three children. Gates are circled, conjunction in blue and disjunction in orange. The trees are authoritative; in the formula labels $\wedge$ binds more tightly than $\vee$, so a label such as $(a\wedge b\vee c)\wedge d$ abbreviates $((a\wedge b)\vee c)\wedge d$.}
\label{fig:nf}
\end{figure}

For a skeleton with $k$ binary connectives and no negation or boundary, the readings are the bracketings of $k+1$ operands, so $|T(w)|=C_k$. The quantity of interest is $|R(w)|$, which is typically far smaller than $|T(w)|$; Figure~\ref{fig:nf} shows the gap for $a\wedge b\vee c\wedge d$.

\section{Admissibility constraints}\label{sec:adm}

Not every derivation tree of a skeleton is a reading an extractor should entertain; two kinds of restriction arise in practice, one syntactic and one semantic, and both fit the framework.

\begin{definition}[Span-local constraint]
A \emph{span-local admissibility constraint} is a predicate $\pi(\mathit{op},i,k,j,w)$ on the root connective $\mathit{op}$ of a subtree, the span $w_i\cdots w_{j-1}$ it covers, and its split point $k$ (for a prefix $\neg$, $k=i$). The admissible readings are $T_\pi(w)$, those all of whose nodes satisfy $\pi$, and the admissible reading classes are $R_\pi(w)=T_\pi(w)/{\equiv_\emptyset}$.
\end{definition}

Two instances are reported by the companion study on clinical eligibility criteria \cite{p01}. \emph{Negation locality} forbids a $\neg$ to cover a span that contains an exception or more than one kind of binary connective; on the pilot skeletons it removed about four percent of the reading classes and no reading that a judge found plausible, and it is adopted. A second constraint, requiring every exception to run to the end of its sentence and attach to the whole preceding material or to the nearest conjunct, removed about fourteen percent of the classes but also four of the twenty plausible readings of the first pilot, because real exceptions attach to intermediate groups, and it is rejected. The point of the span-local form is that, whatever $\pi$ is chosen, the enumeration of Section~\ref{sec:enum} is correct without change, so modeling decisions of this kind are isolated from the algorithm.

The semantic restriction comes from \emph{background axioms}. When a skeleton carries exemplification, as in ``a condition $h$ such as $e_1,\dots,e_r$'', the examples are kinds of the head and contribute hyponymy axioms $e_i\to h$ to $\Gamma$, under which readings that differ only on impossible assignments merge. Since adding axioms can only coarsen the equivalence, that is $\Gamma\subseteq\Gamma'$ implies $\equiv_\Gamma$ refines $\equiv_{\Gamma'}$, enlarging the axiom set can only merge classes, so $|R_\Gamma(w)|\le|R(w)|$; the complexity this adds to deciding ambiguity is taken up in Section~\ref{sec:beyond}.

\section{Enumerating the classes}\label{sec:enum}

The classes can be computed without first materializing the $C_k$ trees, by a dynamic program over spans that stores, for each span, the set of Boolean functions its admissible subreadings realize. For a span $w_i\cdots w_{j-1}$ let
\[
S(i,j)=\{\sem{t}: t\in T_\pi(i,j)\},
\]
computed by $S(i,i{+}1)=\{\sem{w_i}\}$ and, for longer spans, by combining $S(i,k)$ and $S(k{+}1,j)$ at each admissible binary split and $S(i{+}1,j)$ at an admissible prefix negation.

\begin{lemma}[Correctness]\label{lem:enum}
For every span-local constraint $\pi$ and all $i<j$, $S(i,j)=\{\sem{t}: t\in T_\pi(i,j)\}$; in particular, when $w$ has no boundary, $S(0,N)$ is in bijection with the admissible reading classes $R_\pi(w)$.
\end{lemma}
\begin{proof}
By induction on $j-i$. The base case is the single atom. For a longer span, a reading $t\in T_\pi(i,j)$ has a root that is either a prefix $\neg$ over $w_{i+1}\cdots w_{j-1}$ or a binary connective $w_k$ with subtrees $t_1\in T_\pi(i,k)$ and $t_2\in T_\pi(k{+}1,j)$, and the root satisfies $\pi$. Since the semantics is compositional, $\sem{t}$ is $\neg\sem{t'}$ or $\sem{t_1}\mathbin{w_k}\sem{t_2}$, which by the induction hypothesis lie in the set the algorithm forms from $S(i{+}1,j)$ or from $S(i,k)$ and $S(k{+}1,j)$. Conversely every function the algorithm forms is the meaning of such a reading, built from witnesses of its arguments, and the reading is admissible because $\pi$ depends only on the root's span and split. Classes across boundaries are obtained by running the algorithm per sentence and conjoining.
\end{proof}

Representing each function by its truth table of $2^m$ bits, the algorithm uses $O(N^3 s^2 2^m)$ bit operations, where $s=\max_{i,j}|S(i,j)|\le\min(|T(w)|,2^{2^m})$. Classes modulo $\Gamma$ are obtained by merging truth tables that agree on all models of $\Gamma$, or, when $m$ is large, by replacing the truth-table test with the entailment queries $\Gamma\models f\leftrightarrow g$ discharged by an SMT solver \cite{z3,cvc5}.

\section{Counting readings of the positive fragment}\label{sec:count}

Throughout this section a skeleton is \emph{positive}: repetition-free, with connectives in $\{\wedge,\vee\}$ only, no negation and no boundary, so that its operator word is $w=o_1\cdots o_{n-1}\in\{\wedge,\vee\}^{\,n-1}$ over $n$ distinct atoms. The argument has three steps. Every reading computes a positive read-once function, so by Proposition~\ref{prop:readonce} a reading class is a normal form; we characterize which alternating ordered trees are realizable normal forms of $w$ (Lemma~\ref{lem:lca}); and we count them by a recurrence that decomposes a normal form at its root gate (Theorem~\ref{thm:count}). The symmetries of the count, a one-pass ambiguity test, and the worst-case growth follow.

Every reading of a positive skeleton computes a positive read-once function, since each atom occurs once and the operators are $\wedge,\vee$; so by Proposition~\ref{prop:readonce} two readings are logically equivalent exactly when they share a normal form. A normal form on $a_1,\dots,a_n$ is an ordered tree whose internal gates are labeled in $\{\wedge,\vee\}$, strictly alternating between a gate and its gate children, each gate having at least two children.

\begin{lemma}[Realizable normal forms]\label{lem:lca}
An alternating ordered tree as above is the normal form of some reading of $w$ if and only if the label of $\mathrm{lca}(k)$ equals $o_k$ for every $k$ with $1\le k<n$. Distinct such trees are normal forms of logically inequivalent readings.
\end{lemma}
\begin{proof}
In any bracketing of $w$ the binary node that carries the connective $o_k$ is the lowest node whose subtree contains both $a_k$ and $a_{k+1}$, and contracting runs of one operator does not change which gate is the lowest common ancestor of two leaves nor its label; so the normal form of a reading satisfies the condition. Conversely, given an alternating tree satisfying it, binarize each gate left to right; every binary node so created lies between two adjacent atoms whose lowest common ancestor is that gate, so it receives the right connective and the result is a bracketing of $w$ whose contraction is the given tree. The last claim is the uniqueness of the read-once normal form.
\end{proof}

\begin{theorem}[Counting]\label{thm:count}
For a label $\ell\in\{\wedge,\vee\}$ with dual $\ell'$, let $T_\ell(i,j)$ be the number of \emph{realizable} $\ell$-rooted normal forms on the atoms $i,\dots,j$ with $i<j$, that is, alternating trees with root label $\ell$ in which the label of $\mathrm{lca}(k)$ is $o_k$ for every $k$ with $i\le k<j$; and let $M_\ell(c,j)$ be the number of ways to tile the atoms $c,\dots,j$ into at least one consecutive \emph{part}, each part a single atom or a realizable $\ell'$-rooted normal form on its atoms, with $\ell$ at every boundary between parts. Then
\[
\begin{aligned}
T_\ell(i,j)&=\sum_{\substack{i\le k<j\\ o_k=\ell}} p_{\ell'}(i,k)\,M_\ell(k{+}1,j),\\
M_\ell(c,j)&=p_{\ell'}(c,j)+\sum_{\substack{c\le k<j\\ o_k=\ell}} p_{\ell'}(c,k)\,M_\ell(k{+}1,j),
\end{aligned}
\]
where $p_{\ell'}(x,y)=1$ if $x=y$ and $p_{\ell'}(x,y)=T_{\ell'}(x,y)$ otherwise, and for $n\ge2$
\[
|R(w)|=T_\wedge(1,n)+T_\vee(1,n).
\]
\end{theorem}
\begin{proof}
By Lemma~\ref{lem:lca} we count alternating ordered trees in which the label of $\mathrm{lca}(k)$ is $o_k$ for all $k$. Fix the root label $\ell$. The root's children partition $1,\dots,n$ into at least two consecutive blocks; the boundary between two consecutive blocks is a pair $k,k{+}1$ whose lowest common ancestor is the root, so $o_k=\ell$ there, and each block is a single atom or an $\ell'$-rooted tree on its own atoms satisfying the condition for the pairs inside it. The first block covers $i,\dots,k$ for the first chosen boundary $k$, and the rest is a tiling of $k{+}1,\dots,j$ into at least one further block of the same kind, which is what $M_\ell$ counts; every pair $k,k{+}1$ is a boundary at exactly one gate, namely its lowest common ancestor, so the conditions at the different gates are independent and the products count correctly. The root label is determined by the function for $n\ge2$ (the normal form is unique), so the $\wedge$-rooted and $\vee$-rooted families are disjoint and their counts add.
\end{proof}

The recurrence is quadratic in the sub-counts and is evaluated in $O(n^3)$ arithmetic operations. As an independent check we computed the classes of every operator pattern on $n\le 11$ atoms by exhaustive enumeration of all $C_{n-1}$ bracketings and their truth tables, and the recurrence returned the same count in every case; Table~\ref{tab:counts} lists counts for small $n$.

\begin{proposition}[Symmetries]\label{prop:sym}
Let the \emph{composition} of a positive operator word $w$ be the sequence $c(w)=(\ell_1,\dots,\ell_r)$ of the lengths of its maximal runs of equal operators. Then $|R(w)|$ is invariant under reversal, $|R(o_1\cdots o_{n-1})|=|R(o_{n-1}\cdots o_1)|$, and under the $\wedge/\vee$ duality that exchanges the two operators; consequently $|R(w)|$ depends only on $c(w)$.
\end{proposition}
\begin{proof}
Reversing the atom order and mirroring every bracketing is a bijection between the readings of $w$ and those of its reversal, and since $\wedge$ and $\vee$ are commutative a reading and its mirror compute the same function of the atom set, so the bijection preserves logical equivalence and the class counts agree. Exchanging the two operators in a reading replaces its function $f$ by the De Morgan dual $x\mapsto\neg f(\neg x)$, a bijection on Boolean functions that preserves equality, so the class counts of $w$ and of its dual word agree. Finally, a positive word is determined by its composition together with its leading operator, so exactly two words share a composition and they are duals of each other; hence $|R(w)|$ is a function of $c(w)$ alone.
\end{proof}

Summing over all words gives a closed form that calibrates the table.

\begin{proposition}[Schr\"oder sum]\label{prop:schroeder}
For $n\ge2$,\[\sum_{w\in\{\wedge,\vee\}^{n-1}}|R(w)|=2\,s_{n-1},\] where $s_k$ is the $k$-th little Schr\"oder number ($1,3,11,45,197,\dots$), the number of plane trees with $k{+}1$ leaves and no node of out-degree one.
\end{proposition}
\begin{proof}
By Lemma~\ref{lem:lca}, $|R(w)|$ is the number of alternating ordered trees on $a_1,\dots,a_n$ whose gate labels satisfy $\mathrm{label}(\mathrm{lca}(k))=o_k$. Conversely, an ordered tree with every gate of out-degree at least two, with no labels, is realizable for exactly two words: the label of the root gate may be chosen freely, every other gate's label is then forced by alternation, and the word is read off as $o_k=\mathrm{label}(\mathrm{lca}(k))$. The number of such trees is $s_{n-1}$ \cite{stanley}.
\end{proof}

\begin{table}[t]
\centering\footnotesize
\caption{Reading classes of positive skeletons. The raw tree count is the Catalan number $C_{n-1}$; $|R|$ depends only on the composition (Proposition~\ref{prop:sym}). ``alt.'' is the maximally alternating pattern, ``run'' the single-operator pattern; the last column lists selected other compositions.}
\label{tab:counts}
\resizebox{\textwidth}{!}{\begin{tabular}{rrrrl}
\toprule
$n$ & $C_{n-1}$ & $|R|$ alt. & $|R|$ run & selected compositions $\to |R|$\\
\midrule
2 & 1 & 1 & 1 & \\
3 & 2 & 2 & 1 & \\
4 & 5 & 4 & 1 & $(1,2),(2,1)\to3$\\
5 & 14 & 10 & 1 & $(1,2,1)\to8$;\ $(1,1,2),(2,2)\to6$;\ $(1,3),(3,1)\to4$\\
6 & 42 & 24 & 1 & $(1,1,2,1)\to19$;\ $(1,1,1,2)\to18$;\ $(1,2,2)\to15$;\ $(1,3,1)\to13$;\ $(2,1,2)\to9$\\
7 & 132 & 66 & 1 & \\
8 & 429 & 172 & 1 & \\
9 & 1430 & 498 & 1 & \\
10 & 4862 & 1360 & 1 & \\
11 & 16796 & 4066 & 1 & \\
12 & 58786 & 11444 & 1 & \\
13 & 208012 & 34970 & 1 & \\
14 & 742900 & 100520 & 1 & \\
15 & 2674440 & 312066 & 1 & \\
\bottomrule
\end{tabular}}
\end{table}

\begin{theorem}[Ambiguity criterion]\label{thm:crit}
A positive skeleton $w$ is logically ambiguous if and only if its operator word contains both $\wedge$ and $\vee$. Consequently logical ambiguity of the positive fragment is decidable in one pass over $w$, in time $O(n)$.
\end{theorem}
\begin{proof}
If $w$ uses only one operator then all its bracketings are equal by associativity, so $|R(w)|=1$. If $w$ uses both, it has two adjacent positions carrying different operators, that is a factor $x\,\wedge\,y\,\vee\,z$ or $x\,\vee\,y\,\wedge\,z$ on three consecutive atoms $x,y,z$. Choose any bracketing of $w$ that first brackets these three atoms together as a subtree, which is possible by associating outward from that factor. The two internal bracketings of the factor give the non-equivalent subfunctions $(x\wedge y)\vee z$ and $x\wedge(y\vee z)$, which differ at $x=0,y=0,z=1$ (for the dual factor $x\vee y\wedge z$ the subfunctions $(x\vee y)\wedge z$ and $x\vee(y\wedge z)$ differ at $x=1,y=0,z=0$), while the surrounding context, a positive read-once formula in the remaining atoms together with one input fed by the factor, is monotone and depends on that input, since every variable of a read-once formula is essential. A monotone function that depends on an input takes both values at some fixed setting of its other inputs, so substituting two functions that differ on some assignment yields two functions that differ on an extension of that assignment; hence the two whole readings are non-equivalent and $|R(w)|\ge2$. The test is to scan $w$ for one $\wedge$ and one $\vee$.
\end{proof}

The same construction, applied at several places at once, makes the class count exponential.

\begin{corollary}[Exponential lower bound]\label{cor:lb}
For a positive skeleton $w$ on $n$ atoms, $|R(w)|\ge2^{s(w)}$, where $s(w)$ is the largest number of pairwise disjoint three-atom factors of $w$ each using both operators. In particular the maximally alternating word has $s(w)=\lfloor n/3\rfloor$, so its class count is at least $2^{\lfloor n/3\rfloor}$.
\end{corollary}
\begin{proof}
Pick $s=s(w)$ disjoint three-atom factors, each with its two operators different, and fix one bracketing of $w$ that first brackets each factor into a subtree and combines the subtrees and the remaining atoms in any fixed way. Within each factor the two internal bracketings give, as in the proof of Theorem~\ref{thm:crit}, two subfunctions that differ on some assignment; choosing one of the two at each factor independently gives $2^s$ readings. The ambient formula, read-once in the factor-subfunctions and the remaining atoms, is monotone and depends on each of its inputs, so two choices that differ at some factor yield globally inequivalent functions by the argument of Theorem~\ref{thm:crit}; the $2^s$ readings are therefore pairwise inequivalent. In the alternating word every interior atom sits between two different operators, so the triples on atoms $3i{-}2,3i{-}1,3i$ for $1\le i\le\lfloor n/3\rfloor$ are disjoint and each uses both operators.
\end{proof}

\begin{proposition}[Worst case, computational]\label{prop:max}
For every $n\le15$ and every positive operator word on $n$ atoms, $|R(w)|$ is maximized by the maximally alternating word; its count $a(n)$ is the ``alt.'' column of Table~\ref{tab:counts}, beginning $1,2,4,10,24,66,172$ for $n=2,\dots,8$.
\end{proposition}

\begin{theorem}[Identification and growth of the alternating count]\label{thm:growth}
Let $a(n)$ be the number of reading classes of the maximally alternating positive word on $n$ atoms, the ``alt.'' column of Table~\ref{tab:counts}, with $a(1)=1$, and let $A(x)=\sum_{n\ge1}a(n)\,x^n$ and $B(x)=\sum_{n\ \mathrm{odd}}a(n)\,x^n$ be its generating function and the odd part of it. Then
\[
B=x+xB^2+B^3,\qquad A=B+B^2,
\]
so that $A$ is algebraic of degree three, $A^3-(x^2-x+2)A^2+(4x-2x^2)A-2x^2=0$. Consequently $a(2k{+}1)$ is the $k$-th 3-Schr\"oder number (OEIS A027307), $a(2k)$ is its convolution square (A032349), the whole sequence is A137842 shifted by one, $a(n)=\mathrm{A137842}(n{-}1)$, and
\[
a(n)\sim c_n\,\rho^{\,n}\,n^{-3/2},\qquad
\rho=\varphi^{5/2}=\sqrt{\tfrac{11+5\sqrt5}{2}}\approx3.33019,
\]
where $\varphi=(1+\sqrt5)/2$ is the golden ratio and
\[
c_n=\begin{cases}\dfrac{\sqrt{50+30\sqrt5}}{5\sqrt{2\pi}\,\varphi^{5/2}}\approx0.2592,& n\text{ odd},\\[3.5ex]
\dfrac25\sqrt{\dfrac{2(45\sqrt5-100)}{\pi}}\approx0.2519,& n\text{ even};\end{cases}
\]
in particular $\rho$ is strictly below the Catalan base $4$.
\end{theorem}
\begin{proof}
Specializing the recurrence of Theorem~\ref{thm:count} to the alternating word, the count of a span depends only on its length and on the parity of the position of its leading operator. Writing $U,V$ for the generating functions of the numbers of realizable normal forms whose root label equals, respectively differs from, the span's leading operator, $Q=x+U+V$ for that of all of them (the atom alone contributes $x$), and subscripts $e,o$ for the even- and odd-length parts, the recurrence collapses to the finite system
\[
\begin{gathered}
U_e=(x+V_o)Q_o,\quad U_o=(x+V_o)Q_e,\\ V_e=U_eQ_e,\quad V_o=U_eQ_o,\\ Q_e=U_e+V_e,\quad Q_o=x+U_o+V_o,
\end{gathered}
\]
with $A=Q_e+Q_o$ and $B=Q_o$. Eliminating $U_e,U_o,V_e,V_o$ and $Q_e$ gives $Q_o^3+xQ_o^2-Q_o+x=0$, which is $B=x+xB^2+B^3$, together with $Q_e=Q_o^2$; eliminating $B$ from $A=B+B^2$ gives the cubic, of whose three branches the one with $A=x+x^2+O(x^3)$ is the series in question (two branches pass through the origin, since the cubic reduces to $A^2(A-2)$ at $x=0$). The equation $B=x+xB^2+B^3$ is the known functional equation of the generating function $\sum_k A027307(k)\,x^{2k+1}$ of the 3-Schr\"oder numbers, and $A032349$ has generating function $x\,(\sum_k A027307(k)x^k)^2$, which is $Q_e=Q_o^2$ after $x\mapsto x^2$ \cite{oeisA027307,oeisA032349}; this identifies the two parity subsequences and hence the sequence A137842. For the growth, the discriminant of the cubic in $A$ is $-4x^4(x^4+11x^2-1)$, whose real roots are $\pm x_0$ with $x_0^2=(5\sqrt5-11)/2=\varphi^{-5}$ (its other roots, $\pm i\rho$, are larger in modulus); the branch is analytic on $|x|<x_0$ and has a square-root singularity at each of the two dominant points $\pm x_0$, both simple roots of the discriminant. Singularity analysis \cite{flajoletsedgewick} at the conjugate pair gives $a(n)=\rho^n n^{-3/2}(c_++(-1)^nc_-+o(1))$ with $\rho=1/x_0=\varphi^{5/2}$, the contribution of $-x_0$ supplying the factor that alternates with the parity of $n$; the two constants are those of the asymptotics of A027307 and A032349 recorded by Kotesovec \cite{oeisA027307,oeisA032349}, rewritten per atom.
\end{proof}

\begin{remark}
We derived the cubic before learning that the sequence was on record (Section~\ref{sec:related}); the derivation above is self-contained, and the identification with the 3-Schr\"oder numbers adds to our count the closed forms, the P-recursive recurrences, and the lattice-path and colored-ternary-tree interpretations collected in the OEIS entries. As checks, the cubic annihilates the first two hundred coefficients produced by the recurrence of Theorem~\ref{thm:count}, and $a(n)\rho^{-n}n^{3/2}$ evaluated at $n=400$ and $n=399$ agrees with $c_n$ to three digits; the verification is in the accompanying code.
\end{remark}

By Proposition~\ref{prop:max} the alternating word is the maximizer for every $n\le15$, and we conjecture it is the maximizer for all $n$; granting the conjecture, $\rho$ is the worst-case growth base over positive words on $n$ atoms. The separation that matters for autoformalization holds already without it: the raw tree count grows with base $4$ while the alternating class count grows with base $\rho<4$, so on this family the equivalence quotient removes an exponential family of spurious readings and not merely a polynomial factor, and by Corollary~\ref{cor:lb} the surviving class count is itself exponential. An extractor that commits to a bracketing during atomization rather than deferring to the grammar therefore discards such a family of genuine alternatives.

\paragraph{On the maximality conjecture.}
We record what we can prove toward extending Proposition~\ref{prop:max} to all $n$, and why the obvious routes fail. The first tool is a second canonical representative of a class, the left-associated bracketing of its normal form.

\begin{lemma}[Canonical bracketings]\label{lem:canon}
Map each normal form to the bracketing obtained by associating every gate to the left. This is a bijection between $R(w)$ and the bracketings of $w$ in which no node has, as its right child, an internal node carrying the same connective.
\end{lemma}
\begin{proof}
In the left-associated bracketing of a gate with children $c_1,\dots,c_r$ the right child of every node of the gate is one of $c_2,\dots,c_r$, a leaf or a gate of the dual label, so the bracketing has the stated property; conversely, in a bracketing with the property, contracting each maximal chain of left children carrying one connective yields an alternating tree whose left-associated bracketing is the original, and distinct normal forms yield distinct bracketings.
\end{proof}

Decomposing a canonical bracketing of $w\,y$ along the chain of right children from its root, which ends at the last connective $y$, gives a recurrence in the length of the word.

\begin{proposition}[Appending a connective]\label{prop:append}
For a positive word $w=o_1\cdots o_{n-1}$ and a connective $y$, writing $w\,y$ for the word on $n{+}1$ atoms, $w_{\le k}=o_1\cdots o_k$, and $w_{>k}=o_{k+1}\cdots o_{n-1}$,
\[
|R(w\,y)|=|R(w)|+\sum_{\substack{1\le k\le n-1\\ o_k\neq y}}|R(w_{\le k})|\cdot|R(w_{>k})|,
\]
with $|R|$ of the empty word equal to $1$.
\end{proposition}
\begin{proof}
In a canonical bracketing of $w\,y$ follow right children from the root; the chain ends at the node carrying $y$, since the right subtree of any other node is nonempty. If the root carries $y$, the left subtree is a canonical bracketing of $w$, which gives the first term. Otherwise let $o_k$ be the connective carried by the node just above $y$ on the chain; by Lemma~\ref{lem:canon}, $o_k\neq y$, the part of the tree to the left of and including that node is a canonical bracketing of $w_{\le k}$ (its own chain of right children ends at $o_k$), and the left subtree of the node carrying $y$ is a canonical bracketing of $w_{>k}$. The decomposition is reversible.
\end{proof}

For the alternating word the sum runs over the positions of one parity, and Proposition~\ref{prop:append} is then the recurrence behind Theorem~\ref{thm:growth}. It also isolates what a proof of maximality needs. Writing $a(n)$ for the alternating count, the maximality for $n{+}1$ atoms follows from the maximality for $n$ atoms together with the statement that the alternating word also maximizes the increment, which we therefore state as the conjecture to attack.

\begin{conjecture}[Increment maximality]\label{conj:increment}
For every positive word $w$ on $n$ atoms,
\begin{equation}\label{eq:increment}
\max_{y}\ \sum_{\substack{k:\ o_k\neq y}}|R(w_{\le k})|\cdot|R(w_{>k})|\ \le\ a(n{+}1)-a(n).
\end{equation}
\end{conjecture}

By Proposition~\ref{prop:append} and induction on $n$, Conjecture~\ref{conj:increment} implies that the alternating word maximizes $|R|$ for every $n$. We have verified it for every word with $n\le16$. We were not able to prove it, and the reason deserves recording: the count is not monotone under any local change of the word. Flipping one connective can lower it even in the interior of a long run, and refining the composition can lower it: $(2,3)$ gives $10$ classes while its refinements $(2,1,2)$ and $(1,1,3)$ give $9$ and $8$, and $(4,4)$ gives $70$ while $(3,1,4)$ gives $20$, since a singleton run wedged between long runs is nearly inert. Every composition does admit \emph{some} refinement that does not lower the count (verified for $n\le15$), so a monotone path to the alternating word always exists, but no uniform rule selecting it survived our tests, and a Walsh expansion of $|R|$ over the $2^{n-1}$ words shows that the alternating word does not simply align all Fourier signs. A proof will therefore have to be global.

\section{Negation, exceptions, and background axioms}\label{sec:beyond}

\paragraph{Negation and exceptions.}
Adding negation keeps the function read-once over the basis $\{\wedge,\vee,\neg\}$. By De Morgan's laws the negations of a repetition-free formula can be pushed to the leaves, where each atom acquires a fixed sign, and the uniqueness of the positive normal form (Proposition~\ref{prop:readonce}) then applies to the signed formula, so that two readings of a skeleton with negation are logically equivalent if and only if they have the same signed normal form. The counting theory does not transfer as it stands, however: pushing a negation inward exchanges the operators under its scope, and because the scope of a $\neg$ token differs between readings, different readings of one skeleton push down to different positive operator words, so neither Proposition~\ref{prop:sym} nor the recurrence of Theorem~\ref{thm:count} applies to the original word. Counting classes in the presence of negation is therefore a separate problem, which we leave open; the enumeration of Section~\ref{sec:enum} remains exact. The exception operator is covered as well: since $\sem{t_1\EXC t_2}=\sem{t_1}\wedge\neg\sem{t_2}$ and, in a repetition-free skeleton, $t_1$ and $t_2$ share no atom, an $\EXC$ node is a read-once gate like any other, and a skeleton with exceptions remains read-once.

Negation does, however, introduce a kind of ambiguity the positive fragment does not have, namely the \emph{scope of the negation token} itself: the skeleton $\neg\,a\,\wedge\,b$ has the two readings $(\neg a)\wedge b$ and $\neg(a\wedge b)$, which are inequivalent although only one binary operator occurs, so the criterion of Theorem~\ref{thm:crit} is special to the negation-free case. In the enumeration the scope of a prefix $\neg$ is exactly the choice of the span it covers, controlled by the negation-locality constraint of Section~\ref{sec:adm} and admitted by Lemma~\ref{lem:enum} without change.

\paragraph{Background axioms and the complexity of the ambiguity test.}
Under hyponymy axioms $\Gamma$ (Section~\ref{sec:adm}) the counts of Section~\ref{sec:count} are upper bounds, and for $\Gamma=\emptyset$ and a positive skeleton the ambiguity test is Theorem~\ref{thm:crit}. Deciding $\Gamma$-equivalence of two readings is the entailment $\Gamma\models\sem{t_1}\leftrightarrow\sem{t_2}$, which for a set $\Gamma$ of Horn implications and Boolean $t_1,t_2$ is a coNP question in general and is discharged in practice by the solver of Section~\ref{sec:enum}. Deciding $\Gamma$-ambiguity, that is $|R_\Gamma(w)|\ge2$, asks whether two non-$\Gamma$-equivalent readings exist; it is answered by the enumeration of Lemma~\ref{lem:enum} with the truth-table test replaced by entailment, at a cost that inherits the exponential growth of the class count (Corollary~\ref{cor:lb}). Whether $\Gamma$-ambiguity admits a test as cheap as the one-pass criterion of the unaxiomatized positive fragment is open; we expect it does not, because the merging effect of $\Gamma$ is not local to the operator word.

\section{Conclusion}\label{sec:concl}

We have made precise the gap between the Catalan number of bracketings of a connective skeleton and the number of its logically distinct readings. For the positive fragment this gap is governed throughout by the uniqueness of the read-once normal form: the realizable classes are characterized by lowest common ancestors, counted exactly by a two-label recurrence, shown to depend only on the composition of operator runs, and decided for ambiguity in a single pass. The class count is exponential, and the maximally alternating family, which we verify to be the worst case up to fifteen atoms, grows with base $\varphi^{5/2}=\sqrt{(11+5\sqrt5)/2}\approx3.330$, a value obtained exactly from a cubic generating function and lying strictly below the Catalan base $4$. Beyond the positive fragment, normal-form uniqueness continues to decide equivalence for negation, exceptions, and background axioms, though the counting law does not transfer.

Several questions remain, which we state as open problems. (1) Prove that the alternating word maximizes $|R|$ for all $n$, which would upgrade the growth base of Theorem~\ref{thm:growth} to the worst case over all positive words; by Proposition~\ref{prop:append} it suffices to prove Conjecture~\ref{conj:increment}, and the non-monotonicity recorded in Section~\ref{sec:count} rules out local exchange arguments. (2) Give a counting theory for skeletons with negation, where the scope of a $\neg$ token varies between readings and the positive recurrence no longer applies. (3) Decide whether the classes under a given admissibility constraint can be enumerated in output-polynomial time, and characterize the constraints under which enumeration is polynomial, which would connect the combinatorics to the extractor constraints of Section~\ref{sec:adm}. (4) Settle the complexity of $\Gamma$-ambiguity relative to the unaxiomatized case.

\section*{Funding}
This research received no specific grant from any funding agency in the public, commercial, or not-for-profit sectors.

\section*{Conflicts of interest}
The author declares that there is no conflict of interest.

\section*{Data availability}
The code that reproduces every computation in the paper, including the verification of the counting recurrence against exhaustive enumeration, the generating-function derivation, and the checks reported for the maximality conjecture, will be deposited in a public repository with a DOI together with the journal version, and is available from the author on request in the meantime. The companion empirical study is available at \url{https://doi.org/10.5281/zenodo.23600731}.

\section*{Acknowledgments}
The author gratefully acknowledges the College of Computing and Artificial Intelligence and the Research and Innovation Office at the United Arab Emirates University for the institutional resources, including computational facilities, library access, and infrastructural support.

\section*{Declaration of generative AI and AI-assisted technologies in the manuscript preparation process}
During the preparation of this work the author used Claude (Anthropic) to assist with language editing and the refinement of draft passages, with literature searches and the retrieval and verification of bibliographic records, and with code assistance for the script that draws the figure and for the verification scripts that accompany the paper. The mathematical results, their proofs, and the interpretation of the work are the author's own. After using this tool, the author reviewed and edited the content as needed and takes full responsibility for the content of the published article.


\end{document}